\documentclass[iop]{emulateapj}

\shorttitle{Streamer Waves}

\shortauthors{Chen et al.}

\usepackage{color}

\begin{document}

\title{Streamer Waves Driven by Coronal Mass Ejections}

\author{Y. Chen, H. Q. Song, B. Li, L. D. Xia, Z. Wu, H. Fu, {\&} Xing Li\altaffilmark{1}}
\affil{Shandong Provincial Key Laboratory of Optical Astronomy and
Solar-Terrestrial Environment, School of Space Science and
Physics, Shandong University at Weihai, Weihai, China 264209;
yaochen@sdu.edu.cn}

\altaffiltext{1}{On sabbatical leave from Institute of Mathematics
and Physics, Aberystwyth University, UK, SY23 3BZ}

\begin{abstract}
Between July 5th and July 7th 2004, two intriguing fast coronal
mass ejection(CME)-streamer interaction events were recorded by
the Large Angle and Spectrometric Coronagraph (LASCO). At the
beginning of the events, the streamer was pushed aside from their
equilibrium position upon the impact of the rapidly outgoing and
expanding ejecta; then, the streamer structure, mainly the bright
streamer belt, exhibited elegant large scale sinusoidal wavelike
motions. The motions were apparently driven by the restoring
magnetic forces resulting  from the CME impingement, suggestive of
magnetohydrodynamic kink mode propagating outwards along the
plasma sheet of the streamer. The mode is supported collectively
by the streamer-plasma sheet structure and is therefore named ``
streamer wave'' in the present study. With the white light
coronagraph data, we show that the streamer wave has a period of
about 1 hour, a wavelength varying from 2 to 4 solar radii, an
amplitude of about a few tens of solar radii, and a propagating
phase speed in the range 300 to 500 km s$^{-1}$. We also find that
there is a tendancy  for the phase speed to decline with
increasing heliocentric distance. These observations provide good
examples of large scale wave phenomena carried by coronal
structures, and have significance in developing seismological
techniques for diagnosing plasma and magnetic parameters in the
outer corona.
\end{abstract}


\section{INTRODUCTION}
Wave phenomena represent the most fundamental and straightforward
response of a system with plasmas and magnetic fields to
perturbations arising from either interior or exterior. The solar
atmosphere, serving as a good example, is very dynamic { by nature
on}  all relevant temporal-spatial scales, {and} is therefore
expected to be able to support various wave modes with different
observational manifestations. Indeed, with the development of
observational techniques, many types of wave or wavelike phenomena
have been discovered in the solar atmosphere. For instance,
compressible density perturbations moving outwards are detected
inside coronal plumes (Ofman et al., 1997, 1999; DeForest {\&}
Gurman, 1998), propagating longitudinal waves are found in coronal
loops (Berghmans {\&} Clette, 1999), and many other phenomena
driven by nearby solar eruptions, including coronal loop
oscillations (Aschwanden et al., 1999; Nakariakov et al., 1999),
coronal shocks (Sime {\&} Hundhausen, 1987; Sheeley et al., 2000),
and the so-called Moreton (Moreton {\&} Ramsey, 1960) and EIT
waves (Thompson et al., 1998; Wills-Davey {\&} Thompson, 1999),
are observed. Extensive observational and theoretical studies have
been conducted to investigate the nature of these dynamical
phenomena (e.g., Aschwanden, 2004; Nakariacov {\&} Verwichte,
2005; Ofman, 2009; and references therein). These studies,
generally speaking, provide valuable information on the coronal
medium through which the waves propagate.

Helmet streamers are the most conspicuous large-scale quasi-steady
structures extending from the lower to outer corona. In the white
light images observed by a coronagraph, a well developed streamer
is delineated by a sharp brightness boundary. The boundary
separates the streamer from its surroundings. Besides the
boundary, a typical streamer also includes a bunch of closed field
arcades, a streamer cusp, and a high density plasma sheet (also
called the streamer stalk or streamer belt) within which a long
thin current sheet {is} embedded (see, e.g, Pneuman {\&} Kopp,
1971; Suess {\&} Nerney, 2006). On the other hand, coronal mass
ejections (CMEs), representing the largest and most energetic
dynamical process in the corona, may cause global perturbations
with a timescale of minutes to hours. Therefore, close
interactions between CMEs and streamers can frequently occur,
especially during the active phase of a solar cycle when CMEs and
streamers are present at virtually all heliolatitudes. In general,
CME-streamer relevant events can be classified into two groups.
One comprises those events of CMEs originating and erupting from
the streamer interior, like the so-called streamer blowouts
(Howard et al., 1985; Hundhausen 1993) or streamer puffs (Bemporad
et al. 2005). On the other hand,  the events in the second group
result from the streamers being hit on the sides by either CMEs
with expanding structures or by CME-driven disturbances like shock
waves. The collision may cause apparent deflections or kinks of
streamer rays tracing the passage of CME disturbances (Sheeley et
al. 2000). In some cases, the collision may have triggered
reconnections across the streamer current sheet as indicated by
the observed streamer disconnection (e.g., Bemporad, et al.,
2008), the release of plasma blobs along the streamer stalk, or
the formation of streamer in/out pairs (e.g., Sheeley {\&} Wang,
2007).

Given the fundamental role played by wave excitations  in a
disturbed plasma-magnetic field system, one natural question
arises, can the streamer respond in the form of observable waves
or wavelike motions to a strong impact from a CME ejecta? If yes,
what modes are they? In the following text, an answer to the above
questions will be provided with two observational examples of
streamer wavy motions driven by CMEs. Their overall details as
revealed from the white-light coronagraph data will be described
in Section 2. In the 3rd section we present our data manipulation
method to extract the profile of the wavy motion, and give the
resultant physical analysis on one of the two events in the 4th
section. In Section 5, we discuss briefly on the CME-streamer
sources and the other observational event. The final section
presents our conclusions and discussion.

\section{LASCO data: white light and running difference images}

From July 3rd to July 7th 2004, the Large Angle Spectrometric
Coronagraph (LASCO) on the \emph{Solar and Heliospheric
Observatory (SOHO)} observed a well defined bright streamer in the
southwest quadrant with a position angle (PA) of about
231$^{\circ}$. The dynamical behavior of this streamer structure
serves as the main subject of this study. In this interval of
streamer observation, totally 16 CME eruptions with various sizes
and sources are recorded by LASCO according to the online CDAW
(Coordinated Data Analysis Workshops) CME catalog. Two of these
events are relevant to our study, which were first observed by
LASCO C2 at 23:06 UT on July 5th and 20:06 UT on July 6th,
respectively. Both events are classified as fast full-halo BA
(Brightness-Asymmetric) CMEs (check the CDAW web for details of
CME classifications, and many other information), both appear to
be brighter in the southern heliosphere, and both seem to
originate from the backside of the Sun as indicated by the
absence of eruptive features on the front side. The linear
speeds of the two eruptions are 1444 and 1307 km s$^{-1}$, according to the CDAW catalog.
From the online CDAW animations of the
two CMEs, one can see that both eruptions produce visible impact
on the streamer dynamics. At the beginning of the events, the
streamer was pushed aside from their initial equilibrium position
upon the impact of the rapidly outgoing and expanding ejecta;
then, the streamer structure, mainly the bright streamer belt,
exhibited apparent large-scale sinusoidal (or snake-like) wavy
motions. The wavy motions of the streamer are the phenomena we put
our focus on in the present study. In the latter event starting
late of July 6th, the wavy feature is much more obvious.
Therefore, in what follows we first conduct a detailed
analysis on this latter event, and then provide a brief discussion
on the earlier one.

\begin{figure}
\epsscale{1.}  \plotone{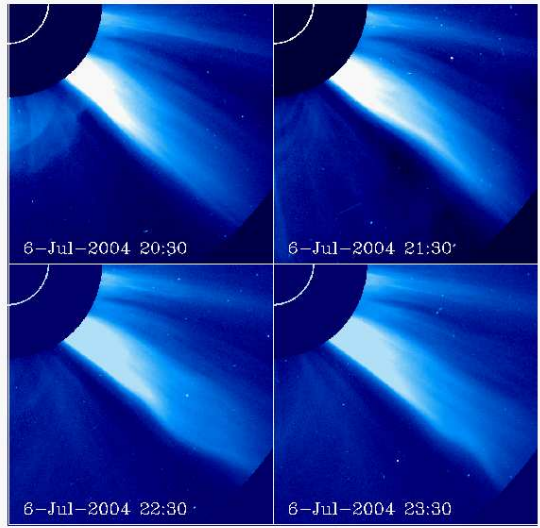} \caption{The wavelike motion of a
streamer stalk observed by LASCO C2 on July 6th, 2004, as an
aftermath of the CME impact. Only the southwest quadrant of the
full field of view of the LASCO C2 observations taken at 20:30,
21:30, 22:30, and 23:30 are shown. The static images at other
moments, relevant running difference images, and animations of the
whole process can be conveniently viewed from the online CDAW CME
database. \label{fig1}}
\end{figure}

To illustrate the overall process of the wavelike motion of the
streamer, in Figure 1 we present four white-light images for
the southwest quadrant of the full field of view (FOV) of the
LASCO C2 observations taken at 20:30, 21:30, 22:30, and 23:30. The
inner white circle represents the size of the Sun, and the black
plate gives the inner occulting disk of LASCO C2. The FOV is
from 2 $R_\odot$ to 7.8 $R_\odot$ for the southwest corner with
the concerned streamer. The CME front enters the C2 FOV at 20:06,
as mentioned. From Figure 1, we see that at 20:30, the
CME front is still in the FOV, which pushes the streamer aside
from its equilibrium position.  By 21:30, the CME front has already
left the FOV {due to} the large outgoing speed of about 1300 km
s$^{-1}$. Comparing the streamer features at the same heights in
the upper two images, we see that the position where the streamer
is strongly deflected gets higher following the CME ejecta, and
the lower part of the streamer starts to bounce back. In the lower
two panels, a sinusoidal motion is clearly seen to propagate
outwards along the streamer stalk. The motion {is also evident}
 from the online animation provided by the CDAW
database. To conduct a more quantitative analysis of the streamer
motion, we need to delineate the wave profiles from the
coronagraph observations. However, it proves difficult to work
directly with the white light images like those shown in Figure 1,
as a result of the inhomogeneous mass distribution of the streamer
structure and the interference of the bright features above the
streamer, especially when the wave amplitude gets smaller with
increasing time and height. Therefore, we examine the running
difference images (RDIs for short) instead, looking for an
alternative approach of plotting the wave profiles. The RDIs,
obtained by subtracting a previous image from the current one, are
extensively used in image manipulations of solar observations to
highlight the region of brightness change between subsequent
exposures.

\begin{figure}
\epsscale{1.}  \plotone{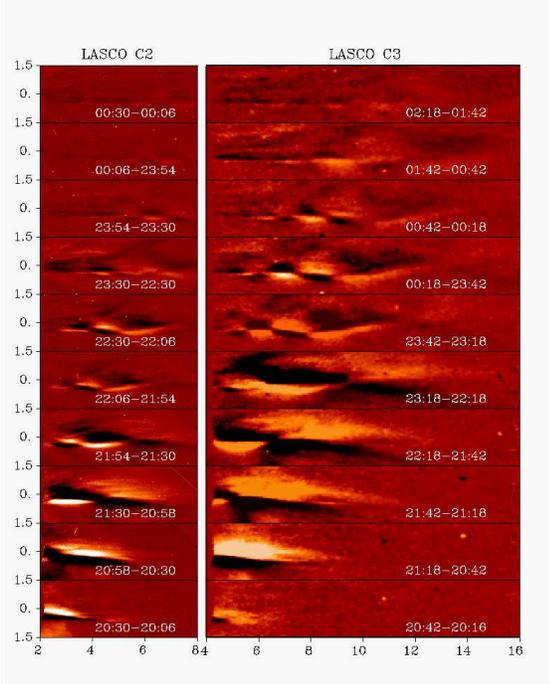} \caption{Two stacks of strips
scissored from the corresponding rotated RDIs from LASCO C2 (left)
and C3 (right). The difference times are given in the figure. The
two vertical sides of the strips are both 1.5 R$_\odot$ away from
the Sun, and the two horizontal sides are 2 (4) and 8 (16) R
$_\odot$ away from the sun for C2(C3)'s observations. See text for
more details. \label{fig2}}
\end{figure}

In Figure 2, we show two stacks of strips scissored from the
corresponding RDIs from LASCO C2 (left) and C3 (right). The
difference times are given in the figure. To get this figure, we
first rotate each RDI counterclockwise by 39$^{\circ}$ to put the
streamer horizontally. Then, we pick out a long strip containing
the streamer. The two vertical sides of the strip are both 1.5
R$_\odot$ away from the sun, and the two horizontal sides are 2
(4) and 8 (16) R $_\odot$ away from the sun for C2(C3)'s
observations. From these RDIs, we notice that most of the bright
features above the streamer that obscure the wave profile now
disappear, and the wavy motions of the streamer, even in the
latter stage of the event when the wave amplitude gets smaller,
are better recognizable. The most prominent feature in the stack
of RDI strips of Figure 2 is given by the interesting bright-dark
and dark-bright pairs. From the definition of image difference,
the presence of such pairs is caused by the displacements of high
density streamer features from bright to dark regions. Therefore,
they are a straightforward manifestation of the propagating
snakelike motion we have observed in the white light images.

Now we take a closer examination of the temporal series of RDI
strips in Figure 2. The bottom three strips, one from C2 and two
from C3, show large areas of bright-upper and dark-lower (BD for
short) regions indicating the deflection of streamer body by the
CME ejection. In the following two ones (C2: 20:58-20:30, C3:
21:42-21:18), a region of opposite pattern with the color distribution
being dark-upper and bright-lower (DB for short) emerges from the
inner part of the corresponding FOV. This is a clear indication
that the streamer starts to bounce backwards from the deflection.
The DB pattern moves outwards to about 4.2 and 6.2 R$_\odot$ in
the FOV of C2 at 21:30 and C3 at 22:18, respectively. In the
strips at 21:54-21:30 for C2 and 23:18-22:18 for C3, the BD
pattern re-emerges indicating the streamer moves again in the same
direction as pushed by the CME,  a result of the overshoot
of the streamer bouncing motion. The change of color pattern from
BD to DB, or vice versa, is continuously observed till the
brightness difference is too weak to discern. Totally, two
to three DB-BD pairs are detected, indicating that there are two
to three observable periods of the streamer wavy perturbation.
From C2 observations, we see that between 20:55 and 21:54 the
streamer accomplishes the first period of wavy motion, and before
22:30, another half period is present. From C3, we see that
between 23:42 and 21:42, two periods of the streamer wavy motion,
i.e., two DB-BD pairs, are formed. We therefore deduce that the
period of the motion is approximately 1 hour. We also realize
that we are actually fortunate to have most inter-exposure
intervals close to one half of the period, i.e., one
half hour. As a result of this coincidence, at a fixed location a
wave crest may be replaced by a wave trough in the subsequent
exposure. This makes the RDIs very suitable for recognizing the
propagating sinusoidal perturbation. There are three RDIs with
inter-exposure intervals being one hour, 23:30-22:30 for C2, and
23:18-22:18 and 01:42-00:42 for C3. It can be seen that the pair
of DB-BD features are not as clear as in the rest.

In the above analysis, we deduce that the period of the streamer
wavy motion is about 1 hour as read from the upward and downward
streamer displacements at the bottom of the LASCO FOV. Apparently,
the restoring force supporting this motion is provided by the
magnetic field of the streamer structure, which comes into play
after the streamer deviates from its equilibrium position upon the
CME impingement. The bouncing motion of the streamer further
excites the outward-propagating sinusoidal perturbations. The
energy received from the initial CME impact is then carried
outwards by the perturbations. Consequently, the bouncing
amplitude of the streamer declines rapidly as observed. The
perturbations are mostly propagating magnetohydrodynamic (MHD)
wave excited by the bouncing motion of the lower part of the
streamer, which is tentatively regarded as the kink mode
collectively supported by the slab configuration of the streamer
plasma sheet (Roberts, 1981; Edwin {\&} Roberts, 1982). More
discussion on the nature of the wave mode will be provided in the
final section of this paper. In the following sections, we shall
call these outward propagating perturbations as streamer wave, and
conduct a more quantitative analysis of their properties.

\section{Method to extract the wave profile}
In this section, we demonstrate the method to extract the wave
profile from RDIs. In general, each RDI contains the information
of two consecutive white light observations, with bright regions
representing where the streamer at the posterior instant of the
difference is, and dark regions giving where the streamer at the
preceding moment was. However, the bright or dark regions usually
are distributed discretely as shown in Figure 2. Therefore, one
needs to make subjective judgement regarding the wave trends when
linking those bright or dark patches together as a whole, and it
is always necessary for one to keep an eye on the associated white
light images to correctly determine the wave profile. In other
words, the deduced wave profile must be consistent with the wavy
motion observed from white light images. In the following, the
method to extract the wave profile is demonstrated using the RDI
strips at 22:30-22:08 for C2 and 23:42-23:18 for C3 which are
shown again in Figure 3 in larger version. The inter-exposure
intervals are both 24 minutes, close to half of the wave period
obtained previously, therefore, both RDIs present well recognized
brightness distribution patterns with DB-BD pairs.

\begin{figure}
\epsscale{1.}  \plotone{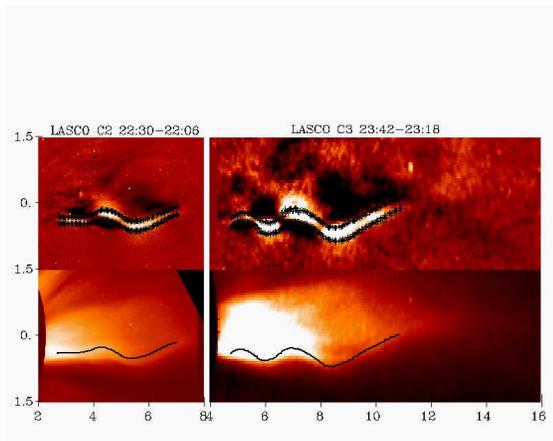}\caption{Examples demonstrating the
method to extract wave profiles from the RDIs obtained by LASCO C2
(left) and C3 (right) observations. The two curves given by plus
signs in the upper panels are obtained by 6 times of delineating
the upper and lower boundaries of the bright patches. The
algebraic average of the two sets of measurements gives the single
curve plotted in the lower panels. \label{fig3}}
\end{figure}

In Figure 3, the two curves given by plus signs in the upper
panels are obtained by 6 times of delineating and linking the
upper and lower boundaries of the bright patches, which are
employed to represent the wave profiles. The algebraic average of
the two sets of measurements gives the single line plotted in the
lower panels where the white light images at 22:30 for C2 and
23:42 for C3 are {also} shown. We see that the wave profile
obtained from boundary delineation in RDIs is basically consistent
with the wavy motion revealed in the white light images. However,
the obtained RDI wave profile does not strictly follow the white
light brightness boundary. The reason is threefold. Firstly, the
bright and dark regions in RDIs indicate the locations where the
brightness or the plasma density changes significantly, while the
brightness boundary in the white light images just reflect the
outer border of the streamer structure. Secondly, the streamer
wave studied here is supported collectively by the streamer
structure, which spreads over a finite range vertically across the
streamer stalk. This range is mainly determined by the width of
the streamer part that is waving, and is apparently different at
different heights. Thirdly, the plasmas are distributed very
non-uniformly across the streamer structure, so plasmas at
different locations may have different weights in supporting the
wavy motion.

As mentioned, there are three RDIs in Figure 2 whose difference
intervals are as large as the wave period, therefore it is not
straightforward to extract the wave profile from these RDIs. For
the RDIs at 23:30-22:30 for C2 and 23:18-22:18 for C3, to yield
the wave profile at the present moment (i.e., 23:30 for C2 and
23:18 for C3) we simply delineate and link the boundary of the
dark patches in the corresponding subsequent RDI, which has an
appropriate exposure interval of 24 minutes. Since the brightness
difference is too weak to be useful at the RDI of 02:18-01:42, we
still delineate and link the bright patches in RDI of 01:42-00:42.
This gives the last wave profile to be shown in Figure 4.

\section{Streamer wave analysis}
Figure 4 shows a collection of streamer wave profiles at various
instants extracted from the rotated RDI strips with the approach
presented in the previous section. The profiles from the C2 (thin curves)
and C3 (thick curves) observations are colored and assembled
into three panels according to the observational times which are
given at the bottom of each panel and colored correspondingly. The
vertical and horizontal scales are both in units of solar radii.
This figure unambiguously confirms the presence of the streamer
wavy motion. The reader can compare any two
consecutive wave profiles with the associated RDI or the white
light images to verify the deduced profiles. With Figure 4, we are
able to conveniently conduct quantitative measurements of wave
properties like the wavelength, the perturbation amplitude, and
the propagation phase speed.

\begin{figure}
\epsscale{1.}  \plotone{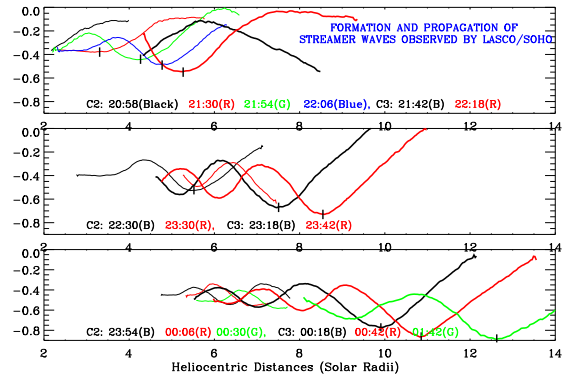}\caption{The streamer wave profiles
at various instants extracted from the rotated RDI strips with the
approach presented in the third section. The profiles from the C2
(thin curves) and C3 (thick curves) observations are colored and
assembled into three panels according to the observational times
which are given at the bottom of each panel and colored
correspondingly. The short vertical lines represent the positions
of Phase P1 at various wave profiles. The vertical and horizontal
scales are both in units of solar radii. \label{fig4}}
\end{figure}

Before doing this, let us first describe some general features as
manifested by the wave profiles. It can be seen that at 20:58, the
downward displacement of the streamer reaches maximum at the
bottom of C2's FOV, this produces a wave trough, which marks the
initiation of the investigated streamer wave. From 20:58 to 21:54,
nearly 1 hour apart, the streamer moves upwards and then downwards
to form a complete wavelength with two wave troughs and a wave
crest. The wave phase associated with the first trough is referred
to as Phase One, or P1 for short. Similarly, the following crest,
the second trough, the second crest, and the third trough are
referred to as P2, P3, P4, and P5, respectively. The heliocentric
distances of these five specified wave phases will be measured to
evaluate the propagation phase speeds.

In Figure 4, the positions of P1 at various wave profiles, when
present, are indicated by black vertical lines. The positions of
other wave phases can be easily read from the associated profiles.
It can be seen that at 22:18, P1 is located at about 5.2
R$_\odot$, which is replaced by the following wave trough P3 one
hour later at 23:18, in agreement with the previous assessment of
the wave period. Similar rough yet consistent estimates can be
carried out using the rest of the wave profiles, for example,
using the anti-correlated red and black thick lines in the lower
two panels.

It is also easy to confirm that there presents two complete
wavelengths spanning from P1 to P5. Hereafter, we refer to the
profile from P1 to P3, i.e., the first wavelength, as W1, and the
profile from P3 to P5 as W2. As seen from the delineated wave
profiles, the complete W1 becomes observable after 21:54 by C2 and
23:18 by C3, and the complete W2 becomes observable after 23:54
for C2 and 00:18 next day for C3. Both the amplitude and
wavelength of W1 get larger during propagation, while the data for
W2 are not sufficient for one to draw similar conclusions. It is
also found that within the same range of heliocentric distances,
the wavelength and amplitude of W2 are smaller than their
counterparts of W1. For example, the W1 wavelength (amplitude) at
23:18 or 23:42 is about 2.5 (0.5) R$_\odot$ larger than the W2
wavelength of 1.7 (0.2) R$_\odot$ at 00:42 within the same range
of 6 - 8 R$_\odot$. The temporal increase of the W1 amplitude may
be attributed to the tendency for the energy flux density to be
conserved during propagation. The decrease of the amplitude from
W1 to W2 is not due to a local damping mechanism of the wave
energy. Instead, it is a result of the convection of the source
energy with the outward propagation of the wave. And, the factor
accounting for the wavelength change will be further discussed
based on the following measurements of the phase speeds.

\begin{figure}
\epsscale{1.}  \plotone{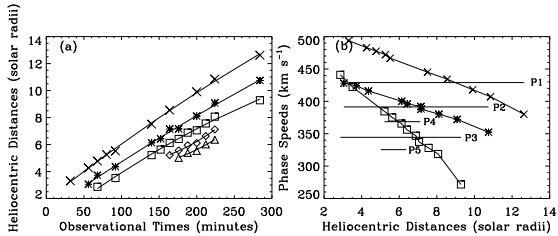} \caption{(a) Heliocentric
distances of the five specific phases P1-P5, represented by
crosses, asterisks, squares, diamonds, and triangles,
respectively. The solid lines are given by a second-order
polynomial fitting to the relevant distance-time profiles. The
horizontal axis is given by the observational times starting from
20:58 UT.  (b) Variation of the linear phase speeds for P1 to P5
with heliocentric distances (horizontal solid lines), and the
phase speeds for P1 (crosses), P2 (asterisks), and P3 (squares)
given by the distance-time polynomial fitting. The relative
uncertainties of the phase distances are estimated to be about
$\pm$10{\%} of the local wavelength, and that of the phase speeds
about $\pm$10{\%} of the plotted values. \label{fig5}}
\end{figure}

Now, we shall focus on the deduction of propagation phase speeds
by measuring and fitting the heliocentric distances of the five
specified phases P1-P5. The obtained distances are shown in Figure
5a, where the crosses, asterisks, squares, diamonds, and triangles
represent the distances of P1 to P5. The solid lines are given by
a second-order polynomial fitting to the relevant distance-time
profiles. The horizontal axis of this figure represents the
observational times starting from 20:58 UT. The increase of the W1
wavelength with time is clearly illustrated by the increase of
distance between P1 and P3 lines. The average phase speeds for P1
to P5 are 429, 391, 344, 369, and 325 km s$^{-1}$, respectively,
which are shown as the 5 horizontal solid lines in Figure 5b,
where the heliocentric distances are used as the abscissa. For
P1-P3, we also show the phase speeds derived by the second-order
polynomial fitting plotted in Figure 5a. The fitted speeds for P4
and P5 are not shown since the accuracy of the polynomial fitting
is greatly reduced by the small number of available distance
measurements.

From the values of the average phase speeds, we see that the
average speed of the preceding phase is faster than that of the
trailing phase, except {that} P4 moves slightly faster than P3 by
about 25 km s$^{-1}$. Such difference of phase speeds is possibly
less than the uncertainties of our measurements, and therefore not
significant. A rough estimate on the uncertainties and relevant
impacts on our conclusions will be given at the end of this
section. From the fitted velocity profiles of P1-P3, we see that
there exists a general trend for the phase speed to decrease with
increasing distance, and again, preceding phases move faster than
trailing ones. If we assume that the period keeps basically
constant during the wave propagation, then the wavelength is
mainly determined by the phase speeds. Thus, the variations of the
phase speeds shown in Figure 5b can provide explanations to
previously mentioned wavelength changes of W1 and from W1 to W2.
To be specific, the result that the difference between the phase
speeds of P1 and P3 gets larger with distance explains the
wavelength increase of W1. The general larger propagation speed of
W1 than W2 explains the positive difference between the two
wavelengths. The large speed variations among various phases at a
fixed location are possibly due to the disturbed state of the
coronal plasmas and magnetic field topologies in the aftermath of
the CME eruption.

Now we proceed to give a rough estimate on the uncertainties of
our measurements on the wavelength and the phase speed. There are
two factors contributing to the errors of our measurements. One
stems from the method we are using to delineate the wave profiles,
and the other is from the determination of the distances of
various phase points in the wave profiles. In the study, the wave
profiles are algebraic averages of two sets of measurements
obtained by delineating and linking the upper and lower boundaries
of the bright (or dark) patches in RDIs. The uncertainties of
determining the wave profiles including the locations of the
crests and troughs should be no larger than one third to a half of
the length of the considered patch, which is about $\pm$10{\%} (or
a total of 20{\%}) of a local wavelength. Once the wave profiles
are plotted, the contribution of the determination of phase point
distances to the total error is not important, as proved by our
practise of determining the phase point distances repeatedly for
several times. The above error to the measurement of the wave
profiles is passed directly to the calculation of the wavelength
and phase speed. Therefore, the errors of the wavelength and phase
speeds given by Figure 5 are estimated to be about $\pm$10{\%} of
the presented values. We see that the phase speed differences
between P3, P4, and P5 are smaller than or close to the relevant
errors, and thus not significant. However, the result of the rough
error analysis does not change our statements regarding the
general variation tendency of the W1 wavelength, and the phase
speeds of the first three phases, as well as the comparisons of W1
and W2 wavelengths, and of P1 to P3 speeds.

The last issue that needs to be addressed in this section is
related to the wave period determined from the above analysis. The
period is about 1 hour, and the interval of the LASCO C2 or C3's
observations are both approximately 30 minutes. Thus, the data are
sampled at roughly the Nyquist rate. This raises the issue of
possible aliasing and incorrect determination of the period if the
actual oscillation period is shorter than 30 minutes (for example,
20 minutes). The issue is addressed from the following two aspects
of argument. Firstly, the concerned imaging areas of the two LASCO
coronagraphs are overlapping between 4 to 8 $R_\odot$, and the
combination of the two sets of observations results in an
effective exposure interval of 12 minutes mostly, as read from the
exposure instants listed in Figures 2 and 4. Secondly, for an
oscillation period as small as, say, 30 minutes, the average phase
speed is 965 km s$^{-1}$ with a wavelength of 2.5 $R_\odot$. As
will be discussed in the discussion section of this paper, this
means an Alfv\'en speed in the slow wind surrounding the plasma
sheet significantly faster than that estimated from previous
relevant theoretical calculations (e.g., Wang et al., 1998; Suess
et al., 1999; Chen {\&} Hu, 2001, 2002; Hu et al., 2003; Li et
al., 2006). According to these calculations, the plasma $\beta$
should be no less than 0.1 in the slow wind regime surrounding the
plasma sheet above the streamer cusp, this yields an Alfv\'en
speed less than 575 km s$^{-1}$ assuming an isothermal temperature
of 1MK for both electrons and protons. Therefore, we conclude that
the value of the deduced wave period is unlikely affected by the
aliasing issue raised above.

\section{Possible CME sources and the earlier event of streamer wave}

\begin{figure}
\epsscale{1.}  \plotone{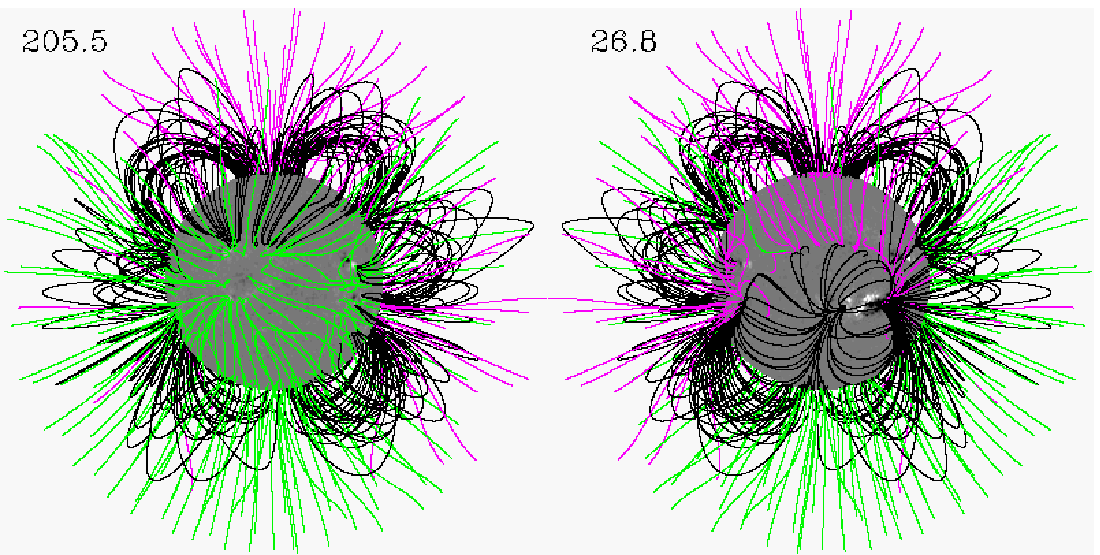} \caption{Coronal magnetic field
topologies obtained from the SSW PFSS model, with central
longitudes being 205.5$^{\circ}$ (left) and 26.8$^{\circ}$
(corresponding to the Carrington times of July 06 20:00, 2004 and
July 20 08:00, 2004). The closed field lines are colored back, and
the open outward (inward) field lines are represented with purple
(green) lines. \label{fig6}}
\end{figure}

\begin{figure}
\epsscale{1.}  \plotone{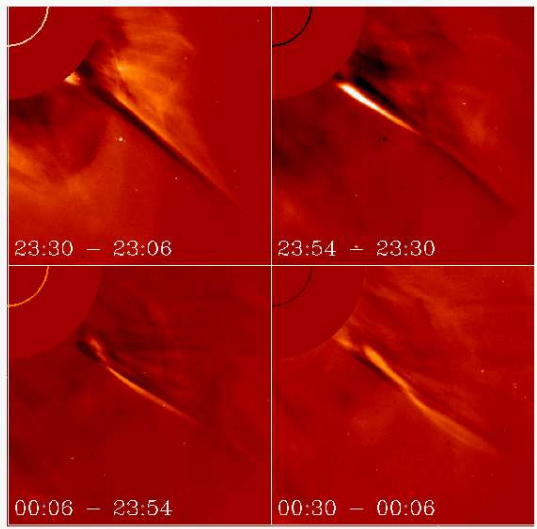} \caption{Four RDIs of the
southwest quadrant of the FOV of LASCO C2 for the CME-streamer
wave event observed on July 5th. \label{fig7}}
\end{figure}

As mentioned at the start of Section 2, the CME eruption that
drives the streamer wavy motion seems to originate from the
backside of the sun. To provide more information on the magnetic
topology of the CME source and the associated streamer, we show
two images of the coronal magnetic fields calculated using the
photospheric fields for Carrington Rotation (CR) 2018 with the
Solar Software (SSW) package PFSS (Potential Field Source Surface,
Schatten et al., 1969). The central meridians of the two images
are taken to be the Carrington longitudes of 205.5$^{\circ}$
(left) and 26.8$^{\circ}$ (right), corresponding to the Carrington
times of July 06 20:00, 2004 and July 20 08:00, 2004. The closed
field lines are colored black, and the open outward (inward) field
lines are represented with purple (green) lines. If assuming the
global magnetic topology does not change significantly during the
CR, we can regard the left image as the front side one and the
right as the backside one at the time of the relevant CME
occurrence. We see that the most probable CME source region is the
active region group in the southeastern quadrant of the backside.
This is consistent with the brightness asymmetric feature of the
eruption. The concerned streamer is also mainly rooted in the
backside, which nominally connects with the suggested CME source
region through a highly inclined loop system. This configuration
allows the CME ejecta to hit directly on the streamer stalk from
the flank without causing any observable disruption of the
streamer.

An earlier CME, first present in the C2 FOV at 23:06 UT on July
5th, is also observed to drive apparent streamer wavy motions. The
overall process of this earlier streamer wave event is presented
in Figure 7 by four RDIs, where the familiar DB-BD features are
observed. The deflection and bouncing of the streamer is evident
from the first and the second images. The third image indicates
that the streamer waves backwards in the direction of the CME
deflection, and the streamer bounces again to the opposite
direction in the last image. It is seen that only one complete
wavelength of the streamer wave is observable. And the streamer
wave feature is not as clear as the one discussed in detail.  A
preliminary evaluation shows that the wave period is also about 1
hour, the wave amplitude, the wavelength, and the propagation
phase speed are about 0.2 R$_\odot$, 2-4 R$_\odot$, and 400 km
s$^{-1}$, respectively.

\section{Conclusions and discussion}
In this paper, we conduct an observational study on the phenomena
of streamer wave, which is excited by the CME impact and
represents one of the largest wave phenomena ever discovered in
the corona. The wave is mostly MHD kink mode propagating outwards
along the thin plasma sheet. The restoring force supporting the
wavy motion is provided by the magnetic field of the streamer
structure, which is generated by the large streamer deflection
upon the CME impact. The energy received from the impact is
carried outwards by the wave perturbation. Consequently, the
amplitude of the wave near the sun declines rapidly with time, and
only a few periods of the wave are observable. The wave period is
estimated to be about 1 hour, the wavelength varies from 2 to 4
R$_\odot$, the wave amplitude is a few tens of solar radii, and
the phase speed is about 300 to 500 km s$^{-1}$. There exists a
general trend for the phase speed to decrease with increasing
heliocentric distance.

Interactions between CME and streamers are frequently observed,
especially during the active phase of solar cycles. Usually, such
interactions result in apparent deflections of interacting
streamers (e.g., Hundhausen et al., 1987; Sime {\&} Hundhausen,
1987; Sheeley et al., 2000). We emphasize that the streamer wavy
motion, reported in the present study, is a direct consequence of
a streamer deflection. Nevertheless, as revealed from a
preliminary overview of the long-term LASCO observations, in only
a very small fraction of the deflection events the streamer
exhibits wavelike phenomena. In other words, most CME-driven
deflections, even very fast and strong, are not followed by a
streamer wavy motion. Therefore, there exist certain strict
conditions for streamer waves to be excited by a CME-streamer
deflection. Two observational features of the July 6 event can
help us evaluate the relevant conditions. Firstly, it is found
that the CME source region lies on the flank side of the closed
loops comprising the streamer, that means the CME does not
originate from beneath the streamer structure, and the ejecta can
collide with the streamer from the flank side. Secondly, the CME
is a fast eruption with a speed of $\ge 1300$~km~s$^{-1}$, which
has two consequences favoring the excitation of the streamer wave.
One is that a faster eruption results in a stronger impingement on
the nearby streamer and a consequent larger deflection of the
streamer structure from its equilibrium position, the other is
that the ejecta moves out of the corona in a relatively short
time, and leaves enough time for the streamer wave to develop.
Otherwise if the eruption is not fast enough, the deflected
streamer may simply moves backwards along with the ejecta, and no
wavy motions result. To observe one example of such a case, one
may check the online LASCO observations of the interaction event
between a CME and a streamer in the northeastern quadrant dated on
July 9th, 2004. Sheeley et al. (2000) also presents LASCO examples
of strong streamer deflection events without accompanying apparent
streamer wavy motions. It should be noted that a more complete
understanding of the excitation conditions of the streamer wave
can only be obtained from observational investigations on much
more similar events and from elaborate theoretical modelling
endeavors.

As mentioned in the introduction section, a well developed typical
streamer consists of the main body, which is a bunch of closed
field arcades confining high density coronal plasmas, and a dense
plasma sheet within which a long thin current sheet is embedded.
The intersection of the closed streamer main body and the open
plasma sheet gives the streamer cusp, which is generally thought
to be below 2 to 2.5 $R_\odot$, very close to the bottom of the
LASCO C2 FOV. After the impact from a CME, the streamer deflects
away from its original equilibrium position. The consequent
restoring motion may excite the wavelike oscillations propagating
along the plasma sheet. Therefore, the geometry supporting the
discussed streamer wave motion can be simplified as a long slender
plasma slab extending to infinity with the lower end attaching to
the streamer cusp which bounces back and forth in a quasi-periodic
manner. The oscillations are observed to be generally transverse
to the nominal direction of the magnetic field. The manifestation
and the geometry of the phenomena are very similar to that of the
well-known kink mode deduced from a slender magnetic slab except
being in a spherical expanding geometry (Roberts, 1981; Edwin {\&}
Roberts, 1982). It is therefore suggested that the wave phenomenon
discussed in this study represents the kink mode, which is, in a
more general sense, a type of fast magnetosonic waves propagating
in an inhomogeneous magnetized plasma environment. It is
interesting to notice that the morphology of the streamer wave
discussed above is very similar to a traditional Chinese dance
named as 'Colored Belt Dance' which is performed by dancers
holding one end of a long belt in color.

An important extension to the coronal wave study is to develop
diagnostic techniques of plasmas and magnetic fields through which
the wave propagates, i.e., to conduct the study of coronal
seismology. In our case, the period and phase speed of the
streamer wave which has been regarded as the propagating kink mode
carried by the thin plasma sheet, if well resolved from
observations, can be used to provide information on magnetic
properties of streamers. Generally speaking, the phase speed for
the wave phenomenon investigated in this study is given by the sum
of two components. The first one is the speed of the solar wind
along the plasma sheet, the medium carrying the mode outwards. The
other is of course the phase speed of the wave mode in the plasma
rest frame. The phase speed for the kink mode under thin plasma
sheet geometry can be tentatively described with available MHD
theory developed for a plasma-slab configuration in cartesian
geometry (Roberts, 1981; Edwin {\&} Roberts, 1982). Substituting
nominal parameters in the slow-wind plasma sheet region above the
streamer cusp into the dispersion relation given by Edwin {\&}
Roberts (1982), we find that the phase speed of the relevant fast
kink body mode $c_k$ is smaller than yet rather close to the
external Alfv\'en speed $v_{Ae}=B_e / \sqrt{\mu_0 n m_p}$, where
$n$ is the proton number density and $m_p$ the proton mass. The
difference between the deduced $c_k$ and $v_{Ae}$ is generally
less than one third of $v_{Ae}$. Therefore, to implement a
preliminary seismological study on the magnetic field strength
$B_e$, we take $v_{Ae}$ to be equal to the kink mode phase speed
$c_k$ estimated from our observations.

Regarding the solar wind conditions in the concerned region, the
readers are referred to relevant observational studies (Sheeley et
al., 1997; Wang et al., 2000; Strachan et al., 2002; Song et al.,
2009) and theoretical modelings (e.g., Wang et al., 1998; Suess et
al., 1999; Chen et al., 2001, 2002; Hu et al., 2003; Li et al.,
2006). In this short discussion, we simply make use of the solar
wind conditions obtained by Chen {\&} Hu (2001). Only two
distances are considered, (1) at 5 $R_\odot$, the solar wind
velocity $v_{sw} = 100$ km s$^{-1}$ and $n = 1\times10^5$
cm$^{-3}$, and (2) at 10 $R_\odot$, $v_{sw} = 200$ km s$^{-1}$ and
$n = 2\times10 ^4$ cm$^{-3}$. With these assumptions, it is
straightforward to deduce $c_k$ and thus $v_{Ae}$ at the plasma
rest frame, and then calculate the value of the magnetic field
strength $B_e$ at the above two distances in the region
surrounding the plasma sheet. Here we only present our
calculations of $B_e$ with the measurements associated with the
second phase point P2, whose speeds are 410 km s$^{-1}$ at
5$R_\odot$ and 360 km s$^{-1}$ at 10 $R_\odot$, as read from
Figure 5. It is found that the magnetic field strength declines
from 0.045 G at 5 $R_\odot$ to 0.01 G at 10 $R_\odot$, indicating
a slightly super-radial expansion of the magnetic flux tube from 5
to 10 $R_\odot$. These values are consistent with the results
given by recent corona and solar wind models (e.g., Li et al.,
2006). A more complete seismological study, together with
sophisticated numerical MHD simulations of CME-streamer
interactions to shed more light on the excitation and propagation
of the waves, should be conducted in future.

\acknowledgements The SOHO/LASCO data used here are produced by a
consortium of the Naval Research Laboratory (USA),
Max-Planck-Institut f\"{u}r Aeronomie (Germany), Laboratoire
d'Astronomie Spatiale (France), and the University of Birmingham
(UK). The CME catalog employed in our study is generated and
maintained at the CDAW Data Center by NASA and The Catholic
University of America in cooperation with the Naval Research
Laboratory. SOHO is a project of international cooperation between
ESA and NASA. This work was supported by grants NNSFC 40774094,
40825014, 40890162, 40904047, NSBRSF G2006CB806304, and A
Foundation for the Author of National Excellent Doctoral
Dissertation of PR China (2007B24). We thank Kai Liu and Chenglong
Shen for their assistance in data manipulations.

\end{document}